\documentclass[12pt,a4paper,dvips]{article}
\usepackage{epsfig,epsf,wrapfig,times,mathptm} 
\renewcommand{\section}[1]{\vspace{6pt} \noindent\mbox{#1} \newline \noindent}
\renewcommand{\subsection}[1]{\vspace{6pt} \noindent\mbox{\underline{#1}} 
\newline \noindent}
\renewcommand{\subsubsection}[1]{\vspace{6pt} \noindent\mbox{\underline{#1}}
\noindent}

\newfont{\sansb}{cmssbx10}
\newfont{\sans}{cmss10}

\begin{document}
{\center RECENT RESULTS ON GAMMA-RAY BURSTS WITH THE BEPPOSAX SATELLITE
\vspace{6pt}\\}
Filippo Frontera \vspace{6pt}\\
{\it Dipartimento di Fisica, Universit\`a di Ferrara, 44100 Ferrara, Italy\\
and \\
Istituto TESRE, CNR, Via P. Gobetti, 101, 40129 Bologna, Italy\\
\vspace{-12pt}\\}
{\center ABSTRACT\\}
Recent results on Gamma-Ray Bursts obtained with the X-ray
Astronomy satellite BeppoSAX are reviewed. Main emphasis is given to the GRBs
simultaneously detected with the Gamma-Ray Burst Monitor (40--700~keV) and
the Wide Field Cameras (1.5--26~keV). These bursts were rapidly localized 
with high
precision, which permitted a prompt pointing of their error boxes
with the Narrow Field Instruments aboard the same satellite. In three cases
of bursts, these prompt observations led to the discovery of an
X-ray afterglow. For two events also an optical transient
was discovered. We review these results and their implications.


\begin{wrapfigure}[23]{r}{7.5cm}
\epsfig{file=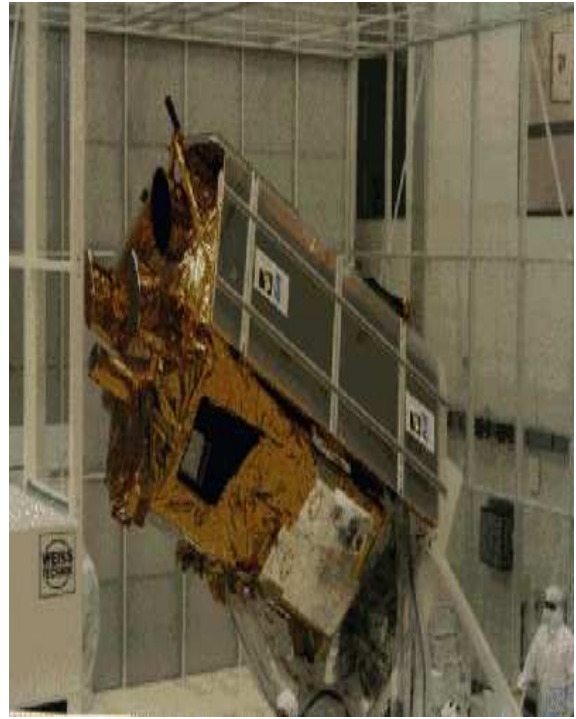,width=7.5cm,height=9.5cm,angle=0}
\caption{The BeppoSAX satellite.}
\end{wrapfigure} 

\setlength{\parindent}{1cm}
\section{INTRODUCTION}
The nature and the origin of celestial Gamma-Ray Bursts (GRB) is recognized
to be one of the major challenges of high energy astrophysics. Before
the launch of BeppoSAX, we knew that they are isotropically distributed
in the sky, while their number versus intensity distribution is not
homogeneous in the space volume accessible to the GRB detectors, with
a deficit of faint bursts (Fishman and Meegan, 1995). A wealth of information
about GRB energy spectra and temporal features has been obtained since their
first discovery by Klebesadel et al. (1973), with several satellite missions
and, in particular, with the {\it Compton Gamma Ray Observatory} (CGRO)
(see, e.g., Preece et al. 1996, Norris et al. 1996, Crider et al. 1997 and
references therein). In spite of these efforts, many scenarios on the nature 
of the bursts are compatible with these data, given the uncertainty on the
sites and distance of these events. Only the detection of an X-ray or optical
counterpart represents the needed break-through to explain the 
GRBs pheonomenon. Thanks to the presence aboard the BeppoSAX satellite (Boella
et al., 1997a) of a Gamma-Ray Burst Monitor (GRBM) with a trigger system in case of
GRB
detection and two Wide Field Cameras (WFC, 1.5 -26 keV, Jager et al., 1997)
capable to quickly provide accurate source positions within a few arcmin
in a field of 40$^\circ \times 40^\circ$, 
we can provide in few hours GRB celestial coordinates and to
point the BeppoSAX Narrow Field Instruments to the 
GRB error box to search for X-ray afterglows. Thus far 5 GRBs have been
simultaneounsly detected
with GRBM-WFC, four of which have been promptly ($<$ 20 hrs) re-observed. In
three cases an X--ray afterglow was clearly discovered  and in
two cases also the optical counterpart was detected by ground-based telescopes.
In this paper we review 
these results and their implications.

\section{THE BEPPOSAX SATELLITE}
SAX (Italian acronym of X-ray Astronomy Satellite) (see Figure 1) is a
major program of the Italian space agency ASI, with
participation of the Dutch space agency NIVR (Boella et al. 1997a).
After the launch the new name of the satellite is BeppoSAX to recall
the Italian physicist Giuseppe (familiarly called Beppo) Occhialini. The main
capability of the mission is to perform spectroscopic and timing studies of
galactic and extragalactic X-ray sources in a broad energy band (0.1-300~keV)
with well balanced instrument performances over the full band. In
the range from 0.1 to 10 keV BeppoSAX can perform spatially resolved studies
of extended sources (e.g., supernova remnants) with 1~arcmin angular
resolution and spectral resolving power E/$\Delta$E in the range from 5
to 10. Also wide sky regions with arcminute angular resolution in the
range from 2 to about 30~keV can be
monitored. Finally  a capability for detecting celestial Gamma-Ray Bursts
(GRBs) is also provided.


\begin{wrapfigure}[25]{r}{7.5cm}
\epsfig{file=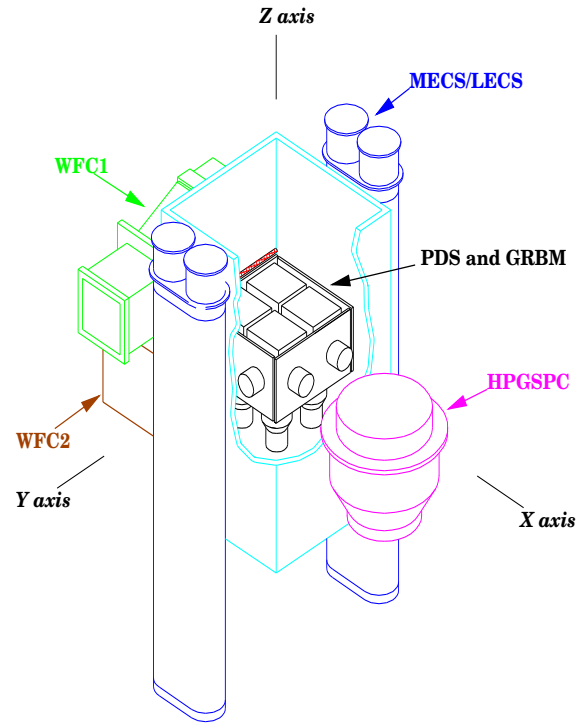,width=7.5cm,height=9.5cm,angle=0}
\caption{Configuration of the BeppoSAX payload. The GRBM consists in the
lateral shields of PDS.}
\end{wrapfigure} 

To achieve all these capabilities, the satellite includes both narrow
and
wide field instruments. The Narrow Field Instruments (NFI) include two
telescopes and two direct viewing detectors. The two telescopes are the
Low Energy Concentrator Spectrometer (LECS), that operates in the 0.1 to
10 keV (Parmar et al. 1997) and the MECS (three units) that operates in the
band from 1.5 to 10 keV (Boella et al. 1997b). Both make use of X-ray
optics. The two direct-viewing detectors are the High Pressure Gas
Scintillator Proportional Counter (HPGSPC), that operates in the band from
3 to 60 keV (Manzo et al. 1997) and the Phoswich Detection System (PDS),
that operates in the 12-300~keV energy band (Frontera et al. 1997).
The NFIs have their axes coaligned. Their field of view ranges from
1.3$^\circ$ (FWHM) for PDS to 30~armcmin for the telescopes.
\\

The wide field instruments include two Wide Field Cameras (WFCs) and
a Gamma-Ray Burst Monitor (GRBM).  Both have their axis orthogonal to
the NFIs. The WFCs are two coded mask proportional counters that look in
opposite directions. They operate in the energy band from 1.5 to 26~keV
with a field of view of 20$^\circ \times 20^\circ$ (FWHM) and imaging
capability with an angular resolution of 3 arcmin. The GRBM is a part of
the PDS experiment.  It is
made of four slabs of CsI(Na) scintillators 10 mm thick, that surround the
core of PDS. The GRBM field of view is almost completely open to the sky.
The surface area of each detection unit is 1100 cm$^2$. The four
scintillators are also used as anticoincidence shields of the PDS.
The GRBM has a dedicated electronics with a trigger system for
fast transient events. Details on
the instrument can be found in the proceedings of this conference (Frontera et 
al. (1997b) and elsewhere
(e.g., Feroci et al. 1997). The GRBM effective energy band is from
40 to 700 keV. A configuration of the BeppoSAX payload is shown in Figure 2.
\\
The satellite was launched on 30 April 1996 from Cape Canaveral (Florida,
USA) with an Atlas-Centaur rocket. Its orbit is almost equatorial 
(inclination = 3.9$^\circ$) at an altitude of about 600 Km. The data are 
stored on board 
and transmitted to the ground after each orbit, when the satellite is visible
from the ground station in Malindi, Kenya. Via a relay
satellite (Intelsat), these data are immediately transmitted to the Operative
Control Center (OCC), located in Rome, Italy. The link with BeppoSAX is also 
used in order to upload telecommands and receive direct telemetry.

\section{BeppoSAX CAPABILITY  FOR GRBs}
The payload configuration described above  makes BeppoSAX highly suitable
for the study of GRBs. From one side we have  two 
detection units of the GRBM with their axis parallel to those of WFCs
with the possibility of monitoring about 5\% of the sky with two
complementary instruments. Counts variations in the WFCs can be recognized
to be due to GRBs and distinguished from, e.g., X-ray burster events, from
their simultaneous detection with the GRBM.
On the other side, the presence on the same satellite of NFIs
with focusing optics affords
the possibility to rapidly perform high sensitivity observations
of GRB error boxes provided by the GRBM/WFC detections.
\\
The GRBM trigger threshold for bursts along the WFC field of view
corresponds to a flux of about 0.6~photons$\,{\rm cm^{-2}\,s^{-1}}$, equivalent
to about $1 \times 10^{-7}\,{\rm erg\,cm^{-2}\,s^{-1}}$. Currently the trigger
system is set for long bursts ($\geq$1~s).
For the triggered bursts, time profiles with high time resolution (down to
0.48~ms) are stored. Continuously we transmit, for each GRBM detection unit,
1~s time profiles in two partially superimposed energy bands (40--700~keV 
and $>$100~keV) and 220 channel energy spectra collected over 128~s.
\\
The WFC sensitivity in 3~s is about $1 \times 10^{-8}\,{\rm erg\,cm^{-2}\,
s^{-1}}$
corresponding to about 0.3 Crab flux unit. The error
radius in the GRB positioning is $\leq$3~arcmin.
\\
Since December 1996, at the BeppoSAX Science Operation Center (SOC) an alert
system is implemented for each simultaneous GRBM/WFC detection
of fast transients events. Following the verification of an actual detection 
of a GRB, a
Target of Opportunity (TOO) pointing of the BeppoSAX NFIs is performed. The
minimum time delay between the first TOO pointing and the initial event can 
be about 5~hrs.

\begin{table}[h]
\vspace{-12pt}
\caption{Gamma-Ray Bursts simultaneously detected with BeppoSAX GRBM and
WFCs}\label{Table}
\begin{center}
\begin{tabular}{ccccc}
\hline\hline
      & TOO time (hrs)  & GRB peak flux & GRB peak flux & Afterglow source \\
Event & from the event  &  40--700~keV  &  2--26~keV    & Flux 1st TOO \\
      &     & erg~cm$^{-2}$~s$^{-1}$ & erg~cm$^{-2}$~s$^{-1}$ & 2--10~keV \\
      &                 &               &         & erg~cm$^{-2}$~s$^{-1}$ \\
\hline
GRB960720 & 1038        & 1.7$\times 10^{-6}$ & 2.5$\times 10^{-8}$ &
QSO 4C~29.29 (?) \\
      &                 &               &             & 1.0$\times 10^{-6}$ \\
\hline
GRB970111 & 16          & 5.6$\times 10^{-6}$ & 1.4$\times 10^{-7}$ &
$\leq$5$\times 10^{-14}$ \\
\hline
GRB970228 &  8          &3.7$\times 10^{-6}$  &1.4$\times 10^{-7}$ &
1SAX J0501.7+1146 \\
     &                  &               &           & 3.$\times 10^{-12}$ \\
\hline
GRB970402 &  8          & 3.2$\times 10^{-7}$ & 1.6$\times 10^{-8}$ &
1SAX J1450.1-6920  \\
     &                  &               &           &2 $\times 10^{-13}$  \\
\hline
GRB970508 &  5.7        & 5.6$\times 10^{-7}$ & 3.5$\times 10^{-8}$ &
1SAX J0653.8+7916  \\
     &                  &               &           & 6 $\times 10^{-13}$  \\

\hline
\hline
\end{tabular}
\end{center}
\end{table}

\section{RESULTS OF GRBM/WFC SIMULTANEOUS DETECTIONS OF GRBs}

Table 1 shows a summary of the simultaneous detections of GRBs obtained thus
far with the BeppoSAX GRBM and WFC instruments. For each of these detections
a TOO observation with the NFIs was performed. The second column of 
Table 1 shows the time  delay between the first TOO observation and the initial
event. As can be seen, the time delay is very long for the first event detected
(GRB960720) and very short for the last event (GRB970508). The long time
delay (about 43~days) for the TOO following the first event is due to the 
fact that the first
simultaneous GRBM/WFC detection occurred during the Science Verification
Phase of the satellite and thus was discovered during the off-line analysis.
The other columns of Table 1 show, for each event, the $\gamma$-ray
(40--700~keV) and X-ray (2--26~keV) peak fluxes achieved during the bursts 
and the result
of the TOO observations. The $\gamma$-ray time profiles of the
last four bursts observed are shown in Figure 3.


\begin{wrapfigure}[26]{r}{7.5cm}
\epsfig{file=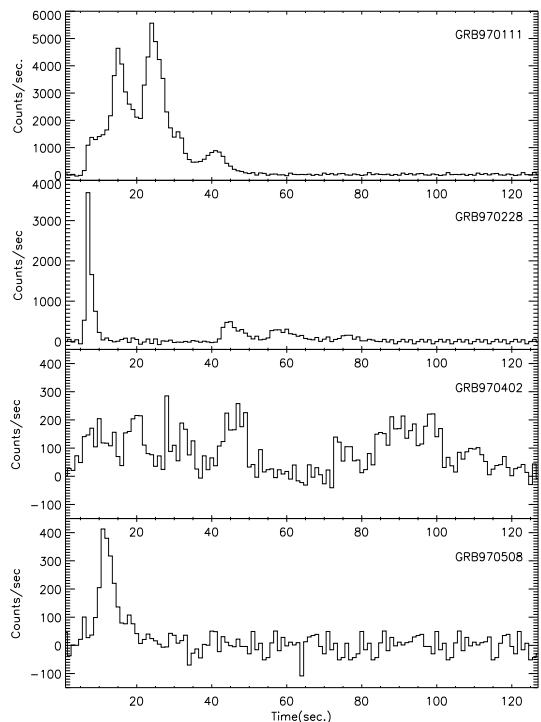,width=7.5cm,height=9.0cm,angle=0}
\caption{Time profiles of the GRBs detected with BeppoSAX GRBM
and WFC, for which a prompt follow-up with the NFIs was performed.}
\end{wrapfigure} 

Now we discuss highlight results obtained for each event.

\subsection{GRB960720}
The event occurred on July 20, 1996 at 11:36:53 UT. Initially its position was
first determined with an error radius of 10 arcmin (Piro et al. 1996).
Successively, with a more refined analysis, the event was  located in an
error circle centred at $\alpha_{2000}\,=\,17^h\, 30^m\,36^s$ and
$\delta_{2000}\,=\,+49^\circ\,05'\,49''$ with 3 arcmin error
radius (in 't Zand et al. 1997). The burst fluence was 2.5$\times
10^{-6} \,{\rm erg\,cm^{-2}}$. A long observation (56~ks) of the 10 arcmin 
error
circle was performed on September 3, 1996 with the NFIs. Results of this
observation
and properties of the burst are reported elsewhere (Piro et al. 1997a).


\begin{wrapfigure}[20]{r}{7.5cm}
\epsfig{file=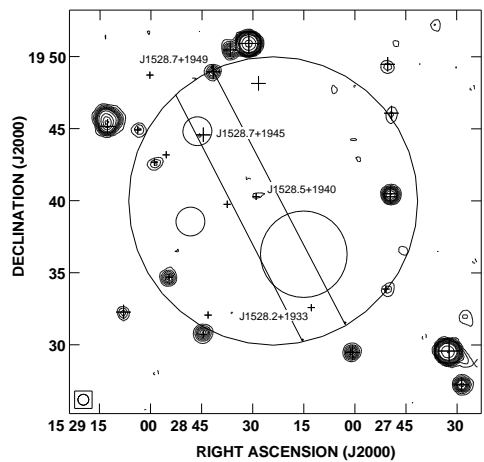,width=7.5cm,height=4.5cm,angle=0}
\caption{WFC preliminary and improved error circles of GRB970111
 along with IPN BATSE/Ulysses error annulus superposed to a radio
continuum image at 1.4~GHz centred on the preliminary GRB centroid position.
Position of the BeppoSAX sources A and B (small open circles) and VLA sources 
are also shown.
(From Frail et al. 1997b.)}
\end{wrapfigure}

A faint X-ray emission in the error box (F(2--10~keV)$\,=\,(1.0\pm0.3)\times
10^{-13}\,{\rm erg\,cm^{-2} \,s^{-1}}$) was found to be consistent with
the direction of the  strong radio
loud quasar 4C~49.29 (f(408~MHz)~=~2~Jy and m$_V$~=~18.8). The ratio of the
measured X-ray flux with the optical one is that  typical
of a
quasar (Maccacaro et al. 1980). Thus it is not immediately possible to relate
the measured X-ray  emission to an X-ray afterglow from GRB960720.

\subsection{GRB970111}

This event was detected on January 11, 1997 at  09:44:00~UT. It was
the strongest of the simultaneous GRBM/WFC detections (see Table 1 for
peak flux) with a total fluence in the 1.5--700~keV energy band of
6.8$\times10^{-5}\,{\rm erg\,cm^{-2}}$. Its $\gamma$-ray time profile is shown
in Figure 3. This burst was the first event localized with
the quick look analysis procedure. Initially its position was
established with a 10 arcmin error radius at the celestial
coordinates $\alpha_{2000}\,=\,15^h \,28^m\,24^s$
and $\delta_{2000}\,=\,+19^\circ\,40'\,00''$ (Costa et al. 1997a). Soon 
after the event, a BeppoSAX
TOO observation of the above error box was approved. The pointing with the NFIs
started 16 hrs after the main event.  Two sources (A and B) in the WFC error
box were discovered (Butler et al. 1997). The ROSAT all-sky survey
data taken in the time period 5--7 August 1991 showed that in the same
error box three sources (1, 2 and 3) were detected (Voges et al. 1997). By
comparing the positions of the ROSAT with those of the SAX sources, it
resulted that the SAX source A was resolved into the ROSAT sources 1 and 2,
while the SAX source B was coincident with the ROSAT source 3 (Voges et al.
1997). The event was also detected with the GRB detector aboard
the Ulysses interplanetary mission and with the BATSE experiment aboard the
CGRO satellite. By using the time delay between these two detections,
Hurley et al. (1997a)  derived an error annulus for the
burst position that reduced the 10 arcmin error circle to a trapezoidal
error box. In this new error box only SAX source A
($\alpha_{2000}\,=\,15^h\,28^m\,46^s$ and $\delta\,=\,+19^\circ\,44'\,50''$)
was contained. The source A appeared a good candidate 
as X-ray counterpart of the GRB970111 afterglow also considering that 
a VLA variable radio source (J1528.7+1945) coincident with the X-ray
position was discovered (Frail et al. 1997a). However, after a more refined
analysis  of the WFC data, the GRB  error box was further reduced  with
a change of the centroid position  (in 't Zand et al. 1997). The new error
circle had a 3 arcmin error radius and was centred at
$\alpha_{2000}\,=\,15^h\,28^m\,15^s$ and $\delta\,=\,+19^\circ\,36'\,18''$.
In the improved error box neither X-ray sources or  variable radio or
optical candidate objects were observed (Frail et al. 1997b, Castro-Tirado
et al. 1997a).
The 3$\sigma$ upper limit to X-ray emission is reported in Table 1.
Figure 4 shows the WFC  (preliminary and refined) error boxes with
the Interplanetery Network (IPN) error annulus and the X-ray (BSAX) and radio
(VLA) source positions.

\subsection{GRB970228}

The gamma-ray burst GRB970228 was the first event for which an
X-ray afterglow was discovered. The BeppoSAX GRBM was triggered by this event 
on February 28, 1997 at 02:58:00 UT (Costa et al. 1997b).


\begin{wrapfigure}[20]{r}{8.0cm}
\epsfig{file=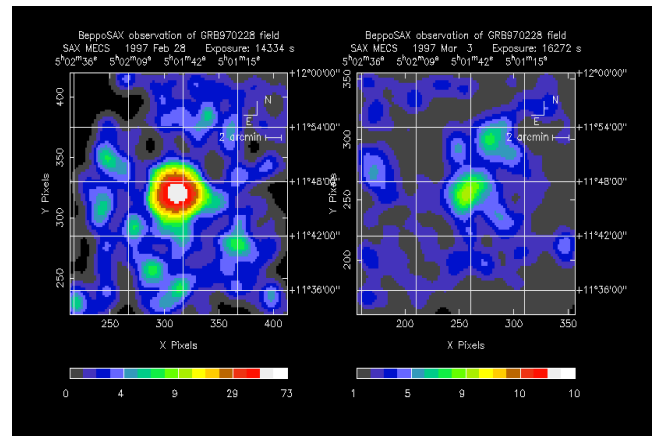,width=8.5cm,height=6.5cm,angle=0}
\caption{False colour-image of the source 1SAX J0501.7+1146, that is
 associated with  GRB970228. (From the paper by Costa et al. 1997d).}
\end{wrapfigure}

Its position was first determined with a 10 arcmin error radius and then
with a radius of 3~arcmin (3$\sigma$) centred at
 $\alpha_{2000}\,=\,05^h01^m57^s$, $\delta_{2000}\,=\,11^\circ46'24''$.
Eight hours after the GRB trigger, from February 28.4681 to February 28.8330
UT, the NFIs were pointed to the WFC error box.
An  X-ray source, SAX J0501.7+1146, was detected (Costa et al 1997c) in the
field of view of both the LECS and MECS telescopes. The source position
($\alpha_{2000}\,=\,05^h01^m44^s$, $\delta_{2000}\,=\,11^\circ46'42''$) was
consistent with the GRB error circle. No previous detection of this
source was obtained with the ROSAT all-sky survey (Boller et al. 1997).
The source was again observed about three days later,
from March 3.7345  to March 4.1174 UT. During this observation the 2-10 keV
source flux had decreased by about a factor 20 while it was not detected
in the 0.1--2~keV energy band. Figure 5 shows the image of the source
in the 2-10~keV band, during the first and the second TOO, obtained with
the MECS telescope.
Following this discovery, searches for radio and optical counterparts of
GRB970228 were started with many  ground-based telescopes located in 
the Northern hemisphere. Galama et al. (1997) first reported the discovery
of an optical transient at a position ($\alpha_{2000}\,=\,05^h01^m46.66^s$,
$\delta_{2000}\,=\,11^\circ46'53.9''$) consistent with both the BeppoSAX WFC
and NFI error circles and with the annuli obtained from the time  delay in
the detection times of the burst  with Ulysses and BeppoSAX  (Hurley et
al. 1997b) and with Ulysses and GGS-Wind experiment (Cline et al. 1997).
Figure 6 shows the sky position of the optical transient (WHT) along with 
various error boxes.
An X-ray observation with the ROSAT HRI instrument of the WFC error box,
performed between March 10.7875 and March 13.32 UT, detected a previously
unknown source, whose centroid position (error circle 10 arcsec) was
consistent with the SAX source and coincident with the optical transient
position within 2 arcsec (Frontera et al. 1997c). This result makes the
association of the optical transient with the SAX source, and thus with
the afterglow of GRB970228, very compelling (Frontera et al. 1997d).
The X-ray afterglow shows very interesting features. The decay curve of
the 2-10~keV flux from 1SAX J0501.7+1146 (see Figure 7) is consistent with a
power law (t$^{-\alpha}$) with $\alpha \,=\,1.33^{+0.13}_{-0.11}$  (Costa
et al. 1997d). When the decay curve is extrapolated backward to the GRB time,
we find that its value is consistent with the average flux of the last three
pulses of the burst (see Figure 3). This fact makes the identification of
the fading X-ray source with GRB970228 very compelling. A power law temporal
decay function with similar slope  is  predicted for the X-ray
flux emitted by a  forward blast wave moving ahead of a 
relativistically
expanding fireball, when it decelerates by ploughing into the surrounding
medium (M\'esz\'aros and Rees 1997, Wijers et al. 1997). In this model
the power law slope is expected  to be independent of the
photon energy.
\\
The spectral evolution of the burst and its
X-ray afterglow show that the nature of the X-ray afterglow emission is
of non thermal origin and similar to the later portion of the burst emission.
(Frontera et al. 1997e). Thus models that assume that GRB phenomenon
is due to cooling of neutron stars can be ruled out and fireball models
are further constrained in their radiation emission processes.


\begin{wrapfigure}[25]{r}{7.5cm}
\epsfig{file=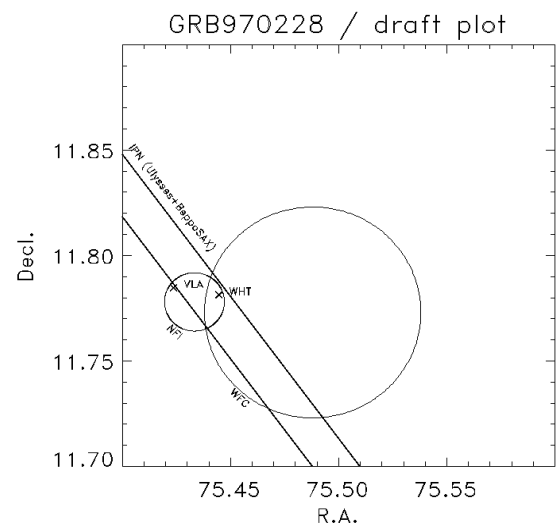,width=7.5cm,height=7.0cm,angle=0}
\caption{Position of the optical transient discovered with William Herschel
Telescope (WHT) and associated with the GRB970228 afterglow. Error boxes of 
the burst obtained with the BeppoSAX WFC and NFIs and error annulus obtained
from the differences in the times the GRB was detected with Ulysses and
BeppoSAX GRBM are also shown.} 
\end{wrapfigure} 

The optical transient associated with GRB970228 was also observed with
the Hubble Space Telescope (HST) 26 and 39 days from the initial event.
At these
epochs the source was no more detectable from ground based telescopes.
A point-like object embedded in an extended nebulosity of about 1 arcsec
extension was seen in both V and I bands (Sahu et al. 1997). While the
point-like object appeared to decline from the first to the second
observation, the emission from the extended source was consistent with
a constant value.
If the extended component is interpreted as the host galaxy of the
optical transient, this result would be in favour of an extragalactic origin
of GRB970228. 
Actually Caraveo et al. (1997), using the above HST observational data, reported
the detection of a proper motion of the optical transient. They estimate
a variation of the transient position by (18$\pm5)\times 10^{-3}\,$arcsec
in 12 days. At the time of writing this paper a further observation
of the optical transient with HST, performed in September 4, has shown 
for the point-like source no proper motion larger than 
100 milli-arcsec/year (Fruchter et al. 1997) disproving the previous claims by 
Caraveo et al. (1997). The above HST observation also confirms the presence of an 
extended optical nebulosity 
superposed to the point-like object (Fruchter et al. 1997). While the intensity of the latter
continues  to decline with a power law  similar to that in the X-ray
band, the intensity of the nebulosity is consistent
with that measured in the March/April observation (Fruchter et al. 1997).

\subsection{GRB970402}
The GRBM detector  was triggered by this
burst (see Table 1 and Figure 3) on April 2, 1997 at 22:19:39 UT (Feroci et
al. 1997). It is the weakest GRB detected with
GRBM and WFC thus far. Its time profile is complex with a long time duration
(about 120~s). The event position was determined with a 3 arcmin error radius
at the celestial coordinates 
 $\alpha_{2000}\,=\,14^h50^m16^s$ and $\delta_{2000}\,=\,-69^\circ19'54''$
(Heise et al. 1997a).
After 8 hrs from the initial event a pointing with the NFIs was performed.
A previously unknown X-ray source, 1SAX J1450.3-6919
($\alpha_{2000}\,=\,14^h\,50^m\,06^s$ and $\delta_{2000}\,=\,-69^\circ\,
20'\,00''$) was discovered. A re-pointing
of BeppoSAX to the source direction after 1.8~days did not show any emission.
For comparison another source, 1SAX J1448.2-6920, that was
in the field, was visible in both pointings (Piro et al.
1997b).

The decay of the source X-ray flux with time is consistent with a power law
with index similar to that of the GRB970228 afterglow
time decay (Nicastro et al. 1997).
Prompt observations of the GRB error box with optical, IR and radio
telescopes did not give any positive result (Castro-Tirado et al. 1997b).


\begin{wrapfigure}[29]{r}{7.5cm}
\epsfig{file=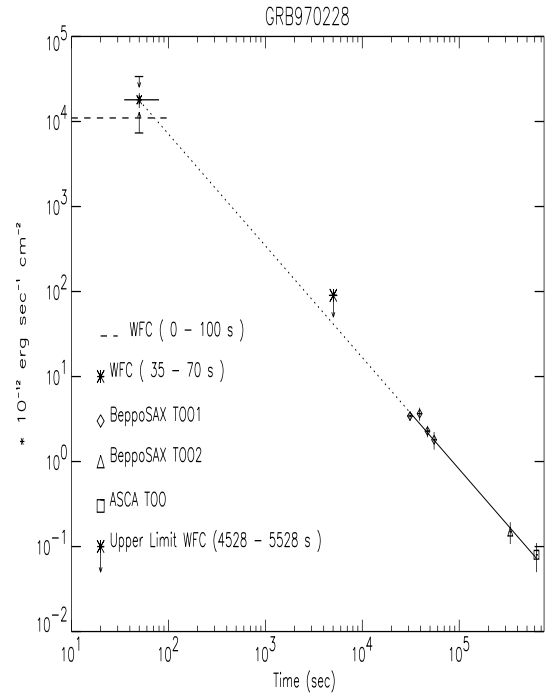,width=7.5cm,height=9.5cm,angle=0}
\caption{Variation of the GRB970228 afterglow with time in the 2--10~keV
range. The solid line shows the fit with a power law (see text).
The extrapolation of this line (dotted)  backwards gives a flux consistent
with the average flux of minor pulses (see fig. 3) of the burst. (From
Costa et al. (1997d)).}
\end{wrapfigure} 

\subsection{GRB970508}

The third X-ray afterglow was observed from the burst GRB970508. This event
was detected by GRBM on May 8, 1997 at 21:41:50 UT. (Costa et al. 1997e).
GRB peak fluxes are given in Table 1. As can be seen, the event is weak with
peak flux 
comparable to that of the April event. However the $\gamma$-ray time profile
is very different (see Figure 3)
with a single pulse and a much shorter duration (about 20~s).
The refined position of the event, obtained with BeppoSAX WFC, is 
$\alpha_{2000}\,=\,06^h\,55^m\,28^s$ and $\delta_{2000}\,=\,-79^\circ\,
17'\,24''$ (Heise et al. 1997b). Following this detection, a TOO observation
of the GRB error box was performed with the NFIs in a shorter time
than in the previous cases (see Table 1). The NFI observation started on
May 9, 1997 at 03:18:00 UT and lasted about 36,000~s. A previously unknown
X-ray source, 1SAX J0653.8+7916, was detected
with LECS and MECS telescopes (Piro et al. 1997c). The source was positioned
at the celestial coordinates $\alpha_{2000}\,=\,06^h\,53^m\,46.7^s$ and
$\delta_{2000}\,=\,-79^\circ\,16'\,02''$ with an error radius of 50''.  In the
same field, another source known from the ROSAT all-sky survey,
1RXS J0653.8+7916, was also detected.
After this first  TOO,  three other observations were performed about
3, 4 and 6 days from the initial event. In the 2-10 keV energy band
the flux decline had a different behaviour
from that of the afterglow source associated with GRB970228. Combining together WFC data
and TOO results, the scenario is of a  source that appears to decline according to a
power law with $\alpha \sim -1.1$ until the time of the first TOO, when a transient event of 
about 10$^5$~s duration superimpose to the above  law (Piro et al. 1997d). 
\\
This peculiar time behaviour of the X-ray afterglow  is accompanied by
a similar complex behaviour of the optical variable (OT J065349+79163)
 discovered by Bond (1997), that has been proposed as probable
optical counterpart of 1SAX J0653.8+7916 (Djorgovski et al. 1997). The
optical transient, after
an indication of decay, showed  a flux rise at the same time as in X-rays. Only  
about 2~days after the
the burst, it showed a definitive decay with a power law function with index
$\alpha\,=\,-1.13\pm 0.04$ (see Figure 8, Pian et al. 1997). No direct detection
of a host
galaxy  has been obtained thus far, in spite of a HST observation of the
optical transient  on June 2, 1997 (Pian et al. 1997). A lower limit to
the host galaxy optical magnitude with R filter is m$_R$~=~24.5. When the 
point-like optical
transient will have further declined, the host galaxy will  be possibly
observable.
\\
A result of primary importance concerning this GRB is the
discovery of redshifted absorption lines in the optical spectrum of
OT J065349+79163 (Metzger et al. 1997). These observations were performed
at the  Keck II 10-m telescope on 11 and 12 May 1997. The source continuum
spectrum is
characterized by a prominent metal absorption line system (mainly Fe II, Mg I and
Mg II) as well as [O II] emission line with wavelengths redshifted by z~=~0.835.
The detection of Mg I absorption suggests the presence of a dense interstellar medium, while
the presence of [O II] emission line suggests stellar formation. Both features suggest the
presence of a galaxy. As a consequence of this
result, the continuum source is either more distant and absorbed by a gas
cloud at this redshift, or is located within the cloud. In any case the
lower limit to the OT J065349+79163  redshift is 0.835.
This is the first determination of the distance  to a GRB.


\begin{wrapfigure}[33]{r}{7.5cm}
\epsfig{file=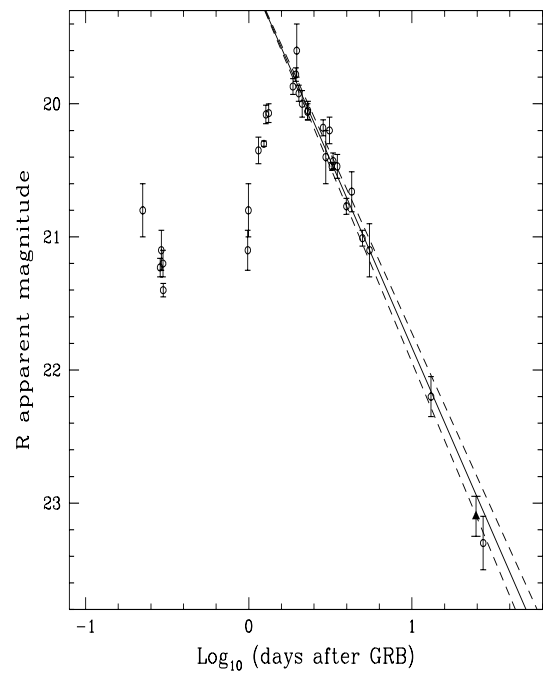,width=7.5cm,height=9.5cm,angle=0}
\caption{Time behaviour of the optical transient associated with
the source 1SAX J0653.8+7916, which in turn is the likely X-ray afterglow
of GRB970508. The continuous line shows the power law fit to the data 
until 10 days after the burst,
while the dashed lines show the 1$\sigma$ confidence interval. The  
apparent magnitude of the OT 1 month later
refers to the HST observation.  
(From Pian et al. 1997.)}
\end{wrapfigure} 

\bigskip

\section{CONCLUSIONS}
The presence of the X-ray afterglow appears an almost common feature of
GRBs. Three of 4 GRBs that were promptly re-observed with the BeppoSAX NFIs
showed X-ray afterglow emission. The burst peak fluxes
cover a large fraction of the log N-log P distribution
derived from the BATSE GRB catalog (Meegan et al. 1996). In addition they
span an order of magnitude in time durations and show different time
profiles. These features do not seem to be relevant for the presence or
absence of X-ray afterglow emission. From only  one burst, GRB970111, we
did not detect any X-ray afterglow  emission after 16 hrs. This could
be related to a faster decline of the X-ray afterglow emission than
in the other bursts or could be intrinsic to the phenomenon itself.
The decline of the X-ray afterglow emission with time is a power law in two
cases and more complex in one case (GRB970508). One possible explanation
of the complex time behaviour of the GRB970508 X-ray and optical afterglow
is the occurrence of an impulsive event with much longer time scale
(about 10$^5$~s) about 3$\times 10^4$~s from the initial event. This
second event could have modified the power law time decay related to the
initial burst (Piro et al. 1997d).
\\
From the optical results, it appears very likely that both GRB970228 and
GRB970508 have extragalactic origin.  As we discussed in section 3.3, the
extragalactic origin is also in agreement with the fireball models
(M\'esz\'aros and Rees 1997, Wijers et al. 1997).
\\
We expect other simultaneous GRBM/WFC detections of GRBs and thus other
detections of GRB afterglows in other bands of the electromagnetic spectrum
in order to clarify several aspects of the GRB phenomenon that are still 
unclear.

\section{ACKNOWLEDGEMENTS}
Many thanks to the BeppoSAX teams who are in charge of
executing the search procedure of simultaneous GRBM-WFC
detections of GRBs. I want to thank Lorenzo Amati for his help in the editing
work of this paper and Elena Pian for a critical reading of the paper.
This research is supported by the Italian Space Agency ASI.

\section{REFERENCES}
\setlength{\parindent}{-5mm}
\begin{list}{}{\topsep 0pt \partopsep 0pt \itemsep 0pt \leftmargin 5mm
\parsep 0pt \itemindent -5mm}
\vspace{-15pt}
\item Boella, G. et al., A\&AS, 122, 299 (1997a).
\item Boella, G. et al., A\&AS, 122, 327 (1997b).
\item Boller, T. Voges, W., Frontera, F. et al., IAU Circular No. 6580
(1997).
\item Bond, H.E., IAU Circular No. 6654 (1997).
\item Butler, R.C., Piro, L., Costa, E. et al., IAU Circular No. 6539 (1997).
\item Caraveo, P.A., Mignani, R.P., Tavani, M. and Bignami, G.F., A\&A,
in press (1997).
\item Castro-Tirado A., et al., IAU Circular No. 6598 (1997a).
\item Castro-Tirado, A. et al., A\&A, submitted (1997b).
\item Cline, T., Butterworth, P.S., Stilwell, D.E., et al., IAU Circular
No. 6593 (1997).
\item Costa, E., Feroci, M., Piro, L., et al., IAU Circular No. 6533 (1997a).
\item Costa, E. et al., IAU Circular No. 6572 (1997b).
\item Costa, E. et al., IAU Circular No. 6576 (1997c).
\item Costa, E., Frontera, F., Heise, J. et al., Nature, 387, 783 (1997d).
\item Costa, E., Feroci, M., Piro, L. et al., IAU Circular No. 6649 (1997e).
\item Crider, A., Liang, E.P., Smith, I.A., et al., ApJ, L39 (1997).
\item Djorgovski, S.G., Metzger, M.R., Kulkarni, S.R., et al., Nature, 387,
      876 (1997).
\item Feroci, M., et al., IAU Circular No. 6610 (1997).
\item Fishman, G.J., and Meegan, C.A., Ann. Rev. of Astr. and Ap., 33, 415
(1995).
\item Frail, D.A., IAU Circular No. 6545 (1997a).
\item Frail, D.A., et al., ApJ Letters, 483, L91 (1997b).
\item Frontera, F. et al., A\&AS, 122, 357 (1997a).
\item Frontera, F. et al., 25th ICRC Conf. Proc., 3, 25 (1997b).
\item Frontera, F. et al., IAU Circular No. 6637 (1997c).
\item Frontera, F., Greiner, J., Costa, E., et al., to be submitted to A\&A
(1997d).
\item Frontera, F. et al., ApJ Letters, submitted (1997e).
\item Fruchter, A. et al., IAU Circular No. 6747 (1997).
\item Hack, F., Hurley, K., Atteia, J.L., et al., AIP Conf. Proc., 307,
359 (1994).
\item Heise, J., in 't Zand, J., Costa, E., and Frontera, F., IAU Circular
      No. 6610 (1997a).
\item Heise, J., in 't Zand, J, Costa, E., et al., IAU Circular No. 6654
(1997b).
\item Hurley, K., Lund, N., Brandt, S., et al., AIP Conf. Proc., 307, 364
(1994).
\item Hurley, K. et al., IAU Circular No. 6571 (1997a).
\item Hurley, K. et al., IAU Circular No. 6594 (1997b).
\item in 't Zand, J., Heise, J., Hoyng, P. et al., IAU Circular No. 6559
(1997).
\item Jager, R. et al., A\&AS, in press (1997).
\item Klebesadel, R.W., Strong, I.B., and Olson, R.A., ApJ, 182, L85 (1973).
\item Maccacaro, T. et al., ApJ, 329, 680 (1980).
\item Meegan, C.A. et al., ApJS, 106, 65 (1996).
\item M\'esz\'aros, P. and Rees, M.J., ApJ, 476, 232 (1997).
\item Metzger, M.R., Djorgovski, S.G., Kulkarni, S.R., et al., Nature, 387,
      878 (1997).
\item Nicastro, L., et al., in preparation (1997).
\item Norris, J.P., Nemiroff, R.J., Bonnell, J.T., et al., ApJ, 459, 393
(1996).
\item Parmar, A. et al., A\&AS, 122, 309 (1997).
\item Pian, E. et al., ApJ (letters), submitted (1997).
\item Piro et al. IAU Circular No. 6467 (1996).
\item Piro, L., Heise, J., Jager, R., et al., A\&A, in press (1997a).
\item Piro, L., Feroci, M., Costa, E., et al., IAU Circular No. 6617 (1997b).
\item Piro, L., Costa, E., Feroci, F., et al., IAU Circular No. 6656 (1997c).
\item Piro et al., in preparation (1997d).
\item Preece, R.D., ApJ, 473, 310 (1996).
\item Sahu, K.C., Livio, M., Petro, L., et al., Nature, 387, 476 (1997).
\item Schaefer, B.E., AIP Conf Proc., 307, 382 (1994).
Van Paradijs, J., Groot, P.J., Galama, T., et al., Nature, 386, 686 (1997).
\item Wijers, R.A.M.J., Rees, M.J., and  M\'esz\'aros, P., MNRAS, 288, L51
(1997).

\end{list}

\end{document}